\documentclass[preprint, nobibnotes, aps, superscriptaddress]{revtex4-1}
\usepackage{amsmath}
\usepackage{amssymb}
\usepackage{hyperref}
\usepackage{physics}
\usepackage{graphicx}
\usepackage{color}
\usepackage{subcaption}

\begin{document}

\title{Inverse Design of Thin--Plate Elastic Wave Devices}

\author{J. R. Capers}
\email{j.capers@exeter.ac.uk}
\affiliation{Centre for Metamaterials Research and Innovation, University of Exeter, Stocker Road, Exeter, EX4 4QL}

\date{\today}

\begin{abstract}
    Motivated by recent advances in the inverse design of electromagnetic materials, we develop two methods for manipulating flexural waves on thin elastic plates. 
    Firstly, we derive a technique for determining plate pinning or mass–loading of a thin plate, designing structures that focus elastic energy, or isolate a region of space from vibrations. 
    Taking inspiration from the adjoint method in electromagnetism, we show how to design graded plates to act as lenses or perform mode shaping.  
    Both of the methods presented are simple, verstile and straightforward to implement, making them useful for designing a wide range of devices for sensing, energy harvesting and vibration isolation.
\end{abstract} 

\maketitle

\section{Introduction}

Controlling mechanical motion is key to many applications such as sensing \cite{Ballantine1997}, energy harvesting \cite{Lee2021, Chaplain2020} and vibration isolation \cite{Zhu2014}.
Elastic materials can be engineered to support a rich variety of wave--shaping behaviours, from mimicking the dispersion relations of condensed matter systems \cite{Torrent2013, Chaplain2023} to exhibiting interesting modal properties \cite{Sabate2021}.
However, few techniques exist to design the properties of elastic systems to manipulate elastic waves in particular ways for particular applications.

While elastic deformations are, in general governed by a set of non-linear tensorial equations \cite{LL7}, when applied to small deformations of a thin plate, the description simplifies to the Kirchhoff-Love equation \cite{Norris1995, Packo2021, Torrent2013, Jose2022}.  
This linear differential equation, a combination of Helmholtz and modified Helmholtz equations, governs the transverse displacement of the plate.
Here we use this simplified description of elastic waves and ask the question `at what points to pin or mass--load a plate in order to manipulate the waves propagating on it in a particular way?'.
Several methods have been developed to answer this question.
Traditional least squares fitting techniques have been utilised to design `metaclusters' of mass--loaded points that scatter flexural waves into particular angles \cite{Packo2021} and analytic methods have been employed to design elastic diffraction gratings \cite{Packo2019}.
For energy harvesting applications, methods utilising adiabatic grading \cite{Chaplain2020, Chaplain2020a} or exploiting disorder \cite{Cao2020} have enabled the design of devices that focus energy to particular points in space.
Machine learning has also been employed to manipulate the properties of edge modes by engineering the unit cell \cite{He2021}.

In addition to pinning or mass loading plates, grading the properties of the plates in space can change how the mechanical waves propagate.
By analogy with results in transformation optics \cite{Leonhardt2010}, coordinate mappings can be used to design invisibility cloaks \cite{Brule2014, Farhat2009, Stenger2012, Farhat2009a, Colquitt2014, Rossi2020, Liu2019, Zareei2017, Darabi2018} for elastic waves on thin plates.
Taking ideas from electromagnetism, graded index lenses for elastic waves have been realised in many ways, including grading plate thickness in space or distributing rods of differing radius in space \cite{Jin2019}.
By grading the height of a thin plate in space one can realise Luneburg, Eaton and Maxwell Fish eye lenses \cite{Climente2014} for elastic waves.
Despite great progress in designing elastic materials to manipulate mechanical waves, there remains the opportunity to develop more versatile inverse design techniques.

In this work, we develop two perturbative techniques for structuring thin plates to manipulate flexural waves.
In Section \ref{sec:dipole} we apply the methods presented in \cite{Capers2021, Capers2022} to design quasi-disordered arrangements of pinned points that can focus flexural waves or isolate regions of space from vibration.
Then, in Section \ref{sec:adjoint}, we take inspiration from the adjoint method in electromagnetism \cite{Lalau-Keraly2013} to grade the properties of thin plates in space to achieve both lensing and mode shaping.
Being both simple and numerically efficient, the techniques we develop may find utility in a range of applications from energy harvesting to sensing.

\section{Quasi-Disordered Arrangements of Pinned Points\label{sec:dipole}}

We begin by developing a method to design quasi-disordered arrangements of pinned points.
Considering a plate of thickness $h$ supporting waves of wavelength $\lambda$, when $h \ll \lambda$ time harmonic flexural waves on the plate obey the following equation \cite{LL7}
\begin{equation}
    \left( \nabla^4 - k_0^4 \right) \phi (\boldsymbol{r}) = 0 ,
    \label{eq:kl}
\end{equation}
where $k_0^4 = h \omega^2 \rho / D$ and for convenience we have introduced the flexural rigidity $D = E h^3 /12(1-\nu^2)$.
The displacement of the plate in the transverse direction is $\phi (\boldsymbol{r})$, with the mechanical properties of the plate expressed through the Young's Modulus $E$ and Poisson's ratio $\nu$.
Under a point excitation at position $\boldsymbol{r'}$, Eq. (\ref{eq:kl}) can be solved in terms of the Green's function, defined as the solution to
\begin{equation}
    \left( \nabla^4 - k_0^4 \right) G(\boldsymbol{r}, \boldsymbol{r'}) = \delta (\boldsymbol{r} - \boldsymbol{r'}) ,
\end{equation}
and given by \cite{Morse1968}
\begin{equation}
    G(\boldsymbol{r}, \boldsymbol{r'}) = \frac{i}{8 k_0^2} \left[ H_0^{(1)} (k_0 |\boldsymbol{r}-\boldsymbol{r'}|) - H_0^{(1)} (i k_0 |\boldsymbol{r}-\boldsymbol{r'}|) \right] ,
    \label{eq:g}
\end{equation}
where $H_0^{(1)} (z)$ is the Hankel function of the first kind.
If many point scatterers e.g. pinning positions or mass loading points are located at positions $\left\{ \boldsymbol{r}_n \right\}$, with scattering strength $\alpha \phi (\boldsymbol{r}_n)$, then the governing equation (\ref{eq:kl}) acquires source terms of the form 
\begin{equation}
    \left( \nabla^4 - k_0^4 \right) \phi(\boldsymbol{r}) = \sum_n \alpha \delta (\boldsymbol{r} - \boldsymbol{r}_n) \phi (\boldsymbol{r}_n) .
    \label{eq:fieldSource}
\end{equation}
This can be directly solved using the Green's function (\ref{eq:g}), with the solution being
\begin{equation}
    \phi (\boldsymbol{r}) = \phi_{\rm inc} (\boldsymbol{r}) + \sum_n \alpha G(\boldsymbol{r}, \boldsymbol{r'}) \phi(\boldsymbol{r}_n) ,
    \label{eq:fieldSln}
\end{equation}
where $\phi_{\rm inc} (\boldsymbol{r})$ is an incident field.
Physically, the point scatterers we are describing could be various arrangements of masses and springs attached to the plate.
Different configurations of masses and springs are described by different models of $\alpha$ \cite{Evans2007, Packo2019}, with the mass--spring system introducing resonances.
In this work, we take the limit of $\alpha \rightarrow \infty$, corresponding to pinning the plate in certain locations i.e. $\phi (\boldsymbol{r}_n) = 0$.
In order to include all multiple scattering effects in the field applied to each scatterer $\phi (\boldsymbol{r}_n)$, a self--consistency condition must be imposed \cite{Foldy1945}.
Taking as the unknown $\alpha \phi (\boldsymbol{r}_n)$, self--consistency results in the matrix equation
\begin{equation}
    \boldsymbol{R} \boldsymbol{\phi}_n = \boldsymbol{\phi}_{i,n} ,
    \label{eq:closure}
\end{equation}
where $\boldsymbol{\phi}_n = \alpha \phi (\boldsymbol{r}_n)$, $\boldsymbol{\phi}_{i,n} = \phi_{\rm inc} (\boldsymbol{r}_n)$ and the matrix $\boldsymbol{R}$ is given by
\begin{equation}
    R_{ij} = \frac{\delta_{ij}}{\alpha} - G(\boldsymbol{r}_i, \boldsymbol{r}_j) .
\end{equation}
Together, the solution to the wave equation (\ref{eq:fieldSln}) and the self--consistency condition (\ref{eq:closure}) allow for the efficient calculation of the displacement of the plate with many pinned points acting as scatterers.

\begin{figure}[h!]
    \centering
    \includegraphics[width=\linewidth]{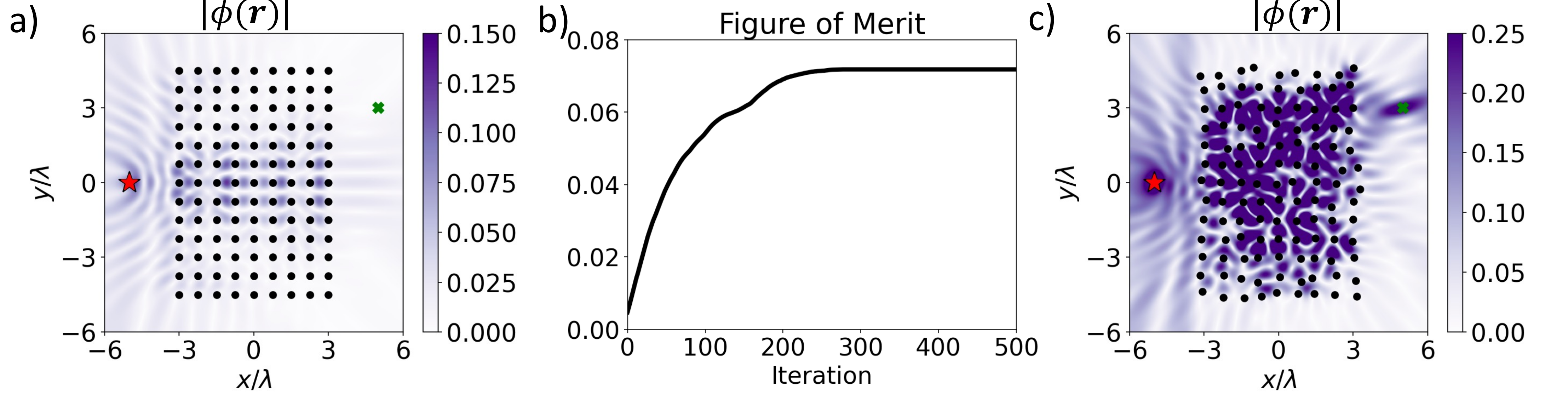}
    \caption{Iteratively designing a structure of pinned points to focus mechanical energy.
    The initial positions of the points are shown in a) as black dots, where the driving point source is shown as a red star.
    We seek to focus energy to the green cross.
    Updating the positions of the scatterers according to the gradient given by Eq. (\ref{eq:analytic_grad}), the figure of merit increase is shown in b).
    The final pinned point configuration and the displacement amplitude is shown in c).}
    \label{fig:source1}
\end{figure}

We now aim to answer the question `where should the plate be pinned to manipulate flexural waves in certain ways?'.
To answer this, we employ the techniques outlined in \cite{Capers2021, Capers2022}, reformulating the key results so they are applicable to the problem of thin elastic plates.
Starting from an initial configuration of pinned points, we will derive an iterative method for increasing a figure of merit.
Asking how a change in the position of a scatterer affects the fields, we perturbatively expand the delta functions in Eq. (\ref{eq:fieldSource})
\begin{equation}
    \delta (\boldsymbol{r}-\boldsymbol{r}_n-\Delta \boldsymbol{r}_n) = \delta (\boldsymbol{r}-\boldsymbol{r}_n) - \Delta \boldsymbol{r}_n \cdot \nabla \delta (\boldsymbol{r}-\boldsymbol{r}_n) + \ldots ,
\end{equation}
as well as the field $\phi \rightarrow \phi + \delta \phi$.
Utilising the properties of the delta function acted upon by a derivative \cite{Gelfand1964}, we collect first order terms to obtain
\begin{equation}
    \delta \phi (\boldsymbol{r}) = - \alpha G(\boldsymbol{r}, \boldsymbol{r}_n) \Delta \boldsymbol{r}_n \cdot \nabla \phi (\boldsymbol{r}_n) .
    \label{eq:fieldChange}
\end{equation}
This expression connects a small change in the position of a scatterer to a small change in the fields, making it useful for finding an analytic expression for the gradient of a figure of merit.
To illustrate this, say we would like to make the displacement large at a particular point $\boldsymbol{r}_\star$.
The figure of merit is therefore
\begin{equation}
    \mathcal{F} = |\phi(\boldsymbol{r}_\star)|^2 .
    \label{eq:fom}
\end{equation}
Expanding this under small changes in the field, to first order we find 
\begin{align}
    \mathcal{F} + \delta \mathcal{F} &= |\phi(\boldsymbol{r}_\star)+\delta \phi(\boldsymbol{r}_\star)|^2 , \\
    &= (\phi^*(\boldsymbol{r}_\star) + \delta \phi^*(\boldsymbol{r}_\star))(\phi(\boldsymbol{r}_\star) + \delta \phi(\boldsymbol{r}_\star)) , \\
    &= |\phi(\boldsymbol{r}_\star)|^2 + 2 {\rm Re}[\phi^*(\boldsymbol{r}_\star) \delta \phi(\boldsymbol{r}_\star)] + \mathcal{O}(\delta \phi^2) .
\end{align}
Substituting Eq. (\ref{eq:fieldChange}) to connect the field change to the change in the position of a scatterer, we find an analytic approximation to the gradient of the figure of merit
\begin{equation}
    \frac{\partial \mathcal{F}}{\partial \boldsymbol{r}_n} = -2{\rm Re} \left[ \phi^* (\boldsymbol{r}_\star) \alpha G(\boldsymbol{r}_\star, \boldsymbol{r}_n) \nabla \phi (\boldsymbol{r}_n) \right] .
    \label{eq:analytic_grad}
\end{equation}
This expression can be combined with simple gradient descent optimisation to find how to update the positions of the scatterers in order to increase the figure of merit Eq. (\ref{eq:fom}) as
\begin{equation}
    \boldsymbol{r}_n^{j+1} = \boldsymbol{r}_n^j + \gamma \frac{\partial \mathcal{F}}{\partial \boldsymbol{r}_n^j} ,
    \label{eq:grad_desc}
\end{equation}
where $j$ is the iteration number and $\gamma$ is a small step size \cite{Sra2012}.
Approaching the optimisation procedure in this way has a number of important benefits.
Firstly, the gradient of the figure of merit with respect to the position of all of the scatterers can be calculated at the same time, making the procedure numerically efficient.
Secondly, while the gradient of the field at each scatterer must be evaluated numerically, the gradient of the figure of merit does not as an analytic expression for it can be derived.
An example of utilising this framework is shown in Figure \ref{fig:source1}.
Considering flexural waves on a steel plate of thickness $h = 1$ mm at $f = 10$ kHz, we take $E = 100$ GPa, $\sigma = 0.3$ and $\rho = 8000$ kg/m$^3$.
These parameter choices give a wavelength of $\lambda \sim$30 mm.
The initial configuration along with the displacement amplitude $|\phi(\boldsymbol{r})|$ is shown in Figure \ref{fig:source1} a).
A point emitter (red star) is placed next to an ordered array of pinned points (black dots).
We aim to increase the displacement amplitude at a target location, indicated by a green cross. 
Using the gradient we have derived Eq. (\ref{eq:analytic_grad}) along with gradient descent optimisation Eq. (\ref{eq:grad_desc}), the scatterers are iteratively moved with the increase in the figure of merit shown in Figure \ref{fig:source1} b).
The final pinned point configuration and displacement amplitude is shown in Figure \ref{fig:source1} c), with a clear peak in amplitude at the desired location.
\begin{figure}[h!]
    \centering
    \includegraphics[width=\linewidth]{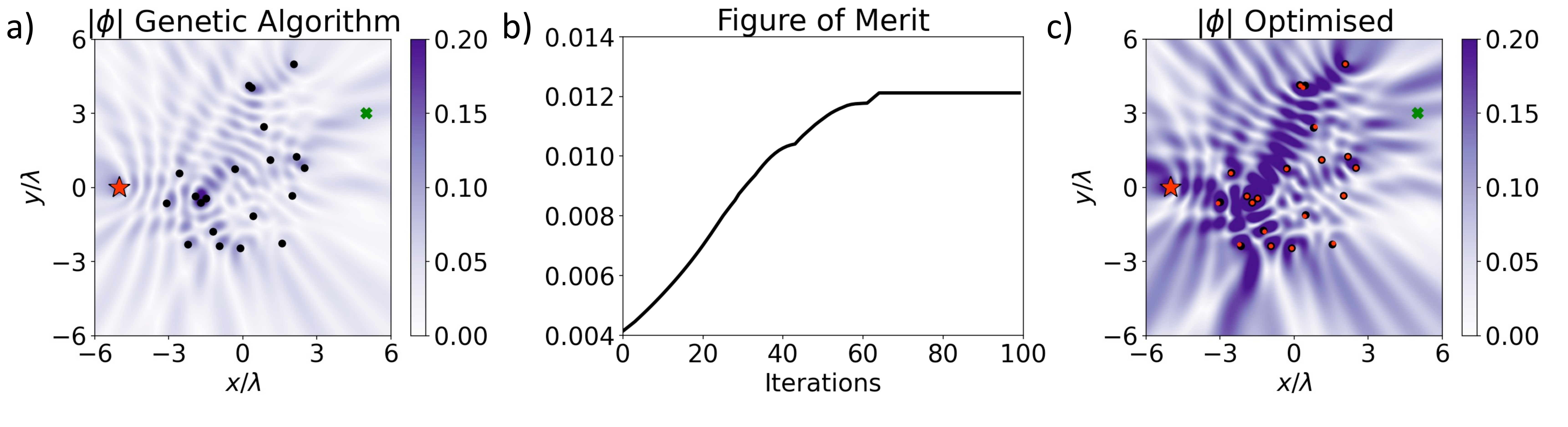}
    \caption{Comparing our optimisation technique to global search methods.
    Seeking to solve the same problem as in Figure \ref{fig:source1}, we a) design an arrangement of 20 pinned points using the differential evolution genetic algorithm.
    To see whether this is a global minima, we use this structure as the starting point of our gradient descent optimisation.  
    The progress of the optimisation is shown in b), with the final structure shown in c).
    Black points indicate the pinned locations given by the genetic algorithm, with red points indicating the optimised pinning locations.}
    \label{fig:go_comp}
\end{figure}
A problem intrinsic to local optimization techniques, including gradient descent, is the likelihood of finding a local rather than a global optima.
However, the problems we are solving have large search spaces as a given number of scatterers can be placed at any point in space.
As a result, while global optimisation techniques such as the genetic algorithm `differential evolution' \cite{Storn1997}, explore much more of the search space a global solution is still not guaranteed.
To illustrate this we consider pinning 20 points to solve the same problem as presented in Figure \ref{fig:source1}.
Employing the differential evolution algorithm with a population size of 20, cross-over probability of CR=0.7 and differential weight parameter of F=0.5, we design the structure shown in Figure \ref{fig:go_comp} a).
To see whether this is a global or local minima, we use this as the starting point for our gradient descent optimisation technique.
The progress of the optimisation and the result are shown in Figure \ref{fig:go_comp} b) and c) respectively.
Our method enhances the figure of merit by a factor of $\sim 3$, meaning the genetic algorithm did not find a global optima.
We conclude that due to the size of the search space, our method is unlikely to find a global optimal, however other methods would also be unable to guarantee this.

\begin{figure}[h!]
    \centering
    \includegraphics[width=\linewidth]{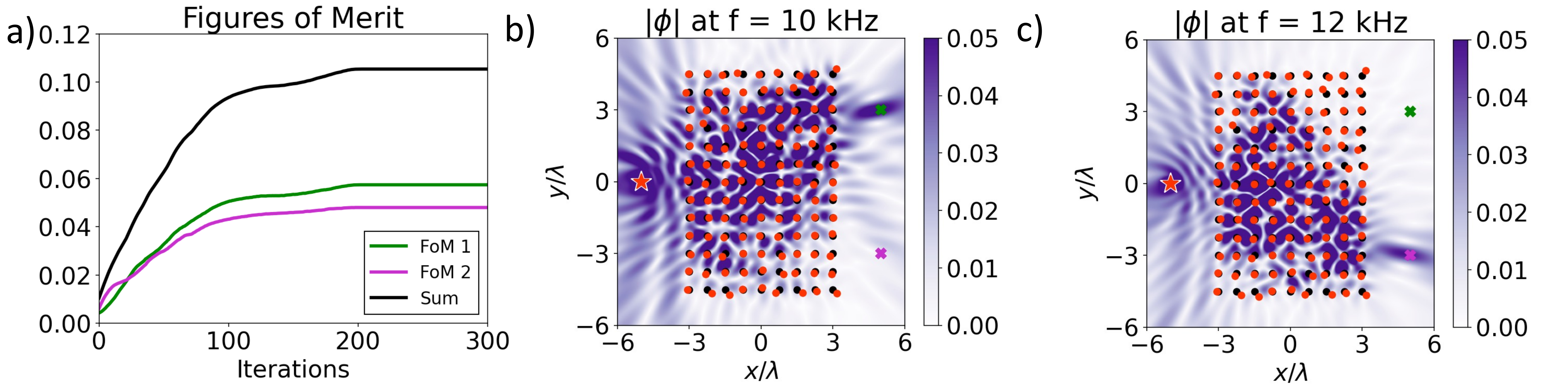}
    \caption{Design of a frequency dependent device.  
    We seek to design a structure that is excited by a point source, shown as a red star, at ($-6 \lambda$, 0) and produces amplitude peaks at different places depending on the frequency of operation.
    Applying our iterative framework, the changes in the figures of merit and their sum is shown in a).
    The displacement amplitude of the final device is shown operating at the two design frequencies, 10 kHz and 12 kHz in b) and c) respectively.
    The target focus point at 10 kHz is shown as a green cross, while the target point at 12 kHz is indicated by a magenta cross.
    Initial pinned point locations are shown as black dots with the optimised pinned locations shown as red dots.}
    \label{fig:source2}
\end{figure}

Taking these ideas further, we can design a structure that couples a point emitter to different locations at different frequencies.
Multiple figures of merit can be optimised by adding them together in a weighted sum, where the weights depend on the current values of each figure of merit \cite{Capers2022}.
Now we seek to simultaneously increase two figures of merit, given by
\begin{align}
    \mathcal{F}_1 &= |\phi (\boldsymbol{r}_1, \omega_1)|^2 & \mathcal{F}_2 &= |\phi (\boldsymbol{r}_2, \omega_2)|^2 . 
\end{align}
For our two frequencies we take 10 kHz and 12 kHz, with the location of the emitter shown as a red star in Figure \ref{fig:source2} and the target locations are green and magenta crosses.
Our figure of merit is therefore 
\begin{equation}
    F = w_1 \mathcal{F}_1 + w_2 \mathcal{F}_2 ,
    \label{eq:fom_sum}
\end{equation}
where the weights are chosen to be $w_i \propto 1/\mathcal{F}_i$ and normalised such that $\sum_i w_i = 1$.
Since we seek to maximise the figure of merit, this choice ensures that if a figure of merit begins small, its weight will be large so the optimisation will favourably enhance it.
Overall, choosing the weights in this way seeks to ensure that both figures of merit in the sum (\ref{eq:fom_sum}) have similar values at the end of the optimisation.
The progress of the optimisation is shown in Figure \ref{fig:source2} a), with the displacement amplitude at 10 kHz and 12 kHz shown in Figure \ref{fig:source2} b) and c) respectively.
The result is a single passive structure that performs different operations at different frequencies.

\begin{figure*}[h!]
    \centering
    \includegraphics[width=\linewidth]{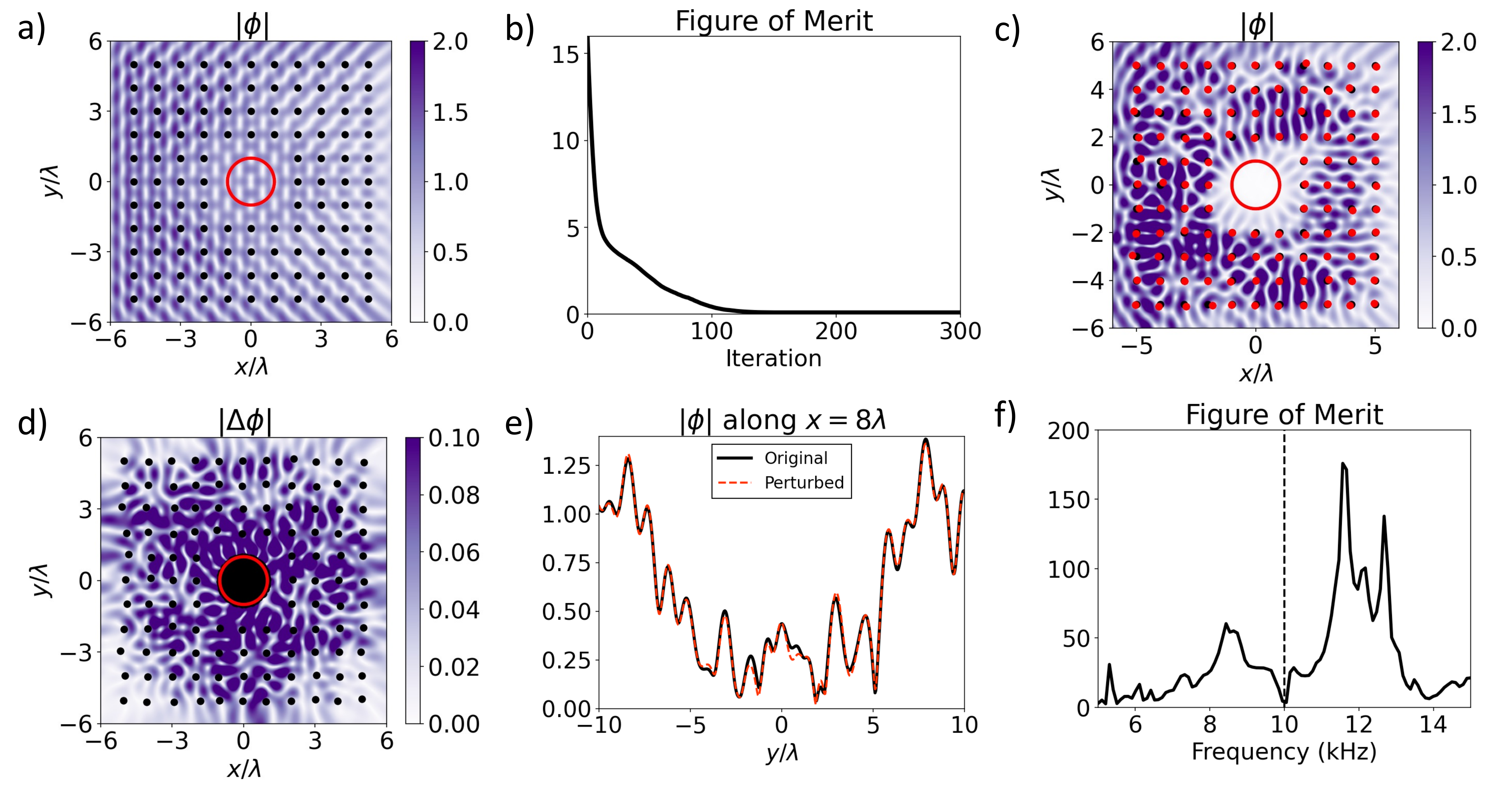}
    \caption{Designing a device that provides vibration isolation for a small region of space.
    An initially regular arrangement of pinned points (black dots) under plane wave incidence has a displacement amplitude shown in a).
    We aim to isolate the region of space indicated by the red circle.
    Updating the pinned locations according to Eq. (\ref{eq:cloakGrad}), the decrease in the figure of merit Eq. (\ref{eq:fom_cloak}) over the optimisation is shown in b).
    The displacement amplitude of the final device is shown in panel c), where the optimised pinning locations are shown as red dots.
    Investigating the behaviour of the device, we pin over 1000 points inside the isolated region then plot d) the change in the field.
    Comparing the the pinned and unpinned case along a cut line, we see e) only small changes in the field.
    Considering f) the bandwidth of the device, we see that the best performance is obtained at the design frequency, with a small bandwidth.}
    \label{fig:cloak}
\end{figure*}

As a final example of the flexibility of this method, we consider trying to isolate a region of space.
We attempt to design a structure that has a region where the flexural waves do not penetrate, so that whatever is placed there has no effect on the scattering as it is completely isolated.
Starting from an ordered structure of pinned points, under plane wave incidence $\phi_{\rm inc} = \exp(ikx)$, we consider a circular region of radius $r = 2\lambda$ centered at the origin, which is to be isolated.
This setup and the displacement amplitude is shown in Figure \ref{fig:cloak} a).
Our figure of merit in this case to be minimised is
\begin{equation}
    \mathcal{F} = \int d^2 \boldsymbol{r'} |\phi(\boldsymbol{r'})|^2 ,
    \label{eq:fom_cloak}
\end{equation}
where the integral is evaluated over the region we seek to isolate.
Expanding this under small changes in the fields, then substituting Eq. (\ref{eq:fieldChange}) we find that the gradient is 
\begin{equation}
    \frac{\partial \mathcal{F}}{\partial \boldsymbol{r}_n} = -2\int d^2 \boldsymbol{r'} {\rm Re} \left[ \phi^* (\boldsymbol{r'}) \alpha G(\boldsymbol{r'}, \boldsymbol{r}_n) \nabla \phi (\boldsymbol{r}_n) \right].
    \label{eq:cloakGrad}
\end{equation}
Evaluating the integral numerically, this gradient can be used design a structure that isolates a small region of space.
The progress of the figure of merit Eq. (\ref{eq:fom_cloak}) over the optimisation is shown in Figure \ref{fig:cloak} b), with the final structure and displacement amplitude shown in c).
While it is evident from the field in Figure \ref{fig:cloak} c) that the region of space that has been targeted is isolated, the performance of the device is considered in Figure \ref{fig:cloak} d)-f).
We consider pinning over 1000 points in the region we have tried to isolate.
If the area is not well isolated, this would significantly affect the scattering.
The difference between the field with the region pinned vs unpinned is shown in Figure \ref{fig:cloak} d).  
The difference in the field is about an order of magnitude lower than the field itself.
Plotting the field along a cut line at $x = 8\lambda$ when the region is pinned and comparing it to the unpinned case, Figure \ref{fig:cloak} e), only small deviations in the field can be observed.
Finally, the bandwidth of the device is considered in Figure \ref{fig:cloak} f) by plotting the figure of merit (\ref{eq:fom_cloak}) as a function of frequency.
At the design frequency of 10 kHz the figure of merit is $\sim 0.05$, although the bandwidth is low at $< 1$ kHz.
Narrow bandwidth is a consequence of the device relying upon many multiple scattering effects to achieve the desired behaviour.
If one instead attached masses on springs, the resulting resonances might be used to engineer a higher bandwidth by grading the properties mass/spring properties in space, as well as their positions.
In other contexts it has been demonstrated that building structures from different species of resonator can enable spectral control i.e. \cite{Chen2018, Nagarajan2018}.
In the context of the method we have presented here, one could attach mass $m_n$ to the plate with a spring of spring constant $k_n$.
Following Torrent et al. the scattering coefficient for the $n^{\rm th}$ scatterer is then
\begin{equation}
    \alpha_n (\omega) = \frac{m_n}{D} \frac{\omega_n^2 \omega^2}{\omega_n^2 - \omega^2} ,
\end{equation}
where $\omega_n = \sqrt{k_n / m_n}$ is the resonant frequency of the mass.
Our optimisation might then be extended to a figure of merit defined over a bandwidth.
For example, to extend the bandwidth of the device presented in Figure \ref{fig:source1}, one might seek to optimise over a given frequency band
\begin{equation}
    \mathcal{F} = \int_{\omega_1}^{\omega_2} d\omega |\phi (\boldsymbol{r}_\star), \omega| .
\end{equation}
Then, as well as moving the locations on the pinned points, one could then find the analytic form of $\partial \mathcal{F} / \partial \omega_n$ and use this to update the resonance frequencies iteratively to attempt to increase the bandwidth.

\section{Grading Plates\label{sec:adjoint}}

\begin{figure}[h!]
    \centering
    \includegraphics[width=\linewidth]{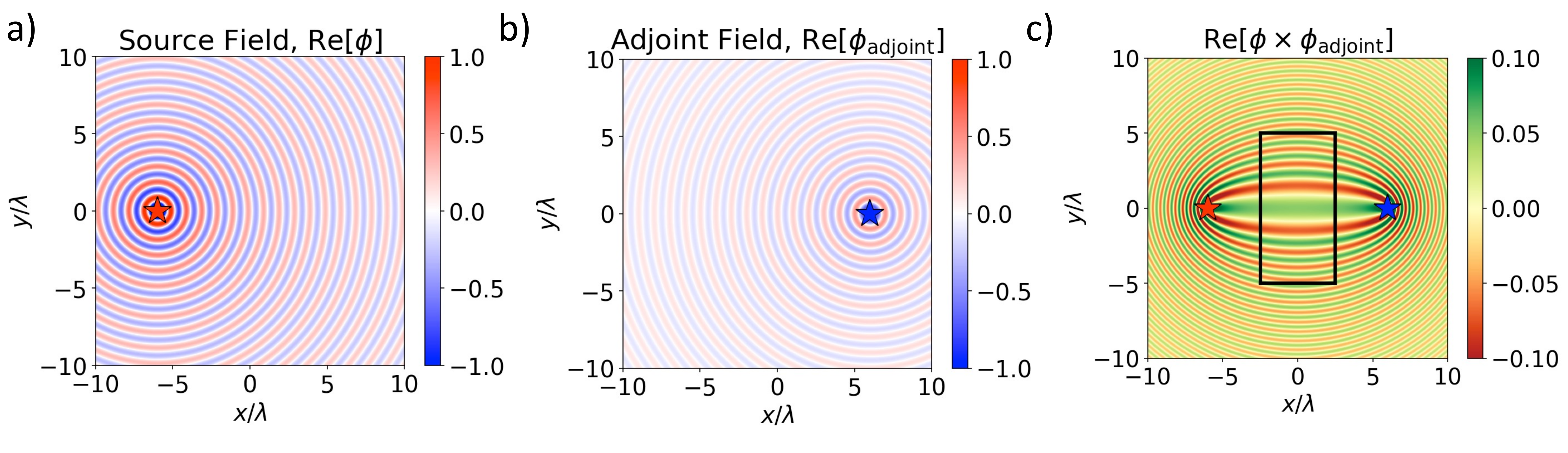}
    \caption{A schematic of how the adjoint method works.
    a) shows the forwards field due to a point source located at the red star.
    The adjoint field b) is then calculated according to Eq. (\ref{eq:adjoint_field}), corresponding to a point source located at the blue star.
    Interfering these two fields we obtain c) how to increase or decrease the mass density to increase the figure of merit.
    Red regions indicate where mass density should be decrease and green regions indicate were it should be increased.
    The region of space we will grade is indicated by the black rectangle.}
    \label{fig:adjoint_method}
\end{figure}

\begin{figure*}[h!]
    \centering
    \includegraphics[width=\linewidth]{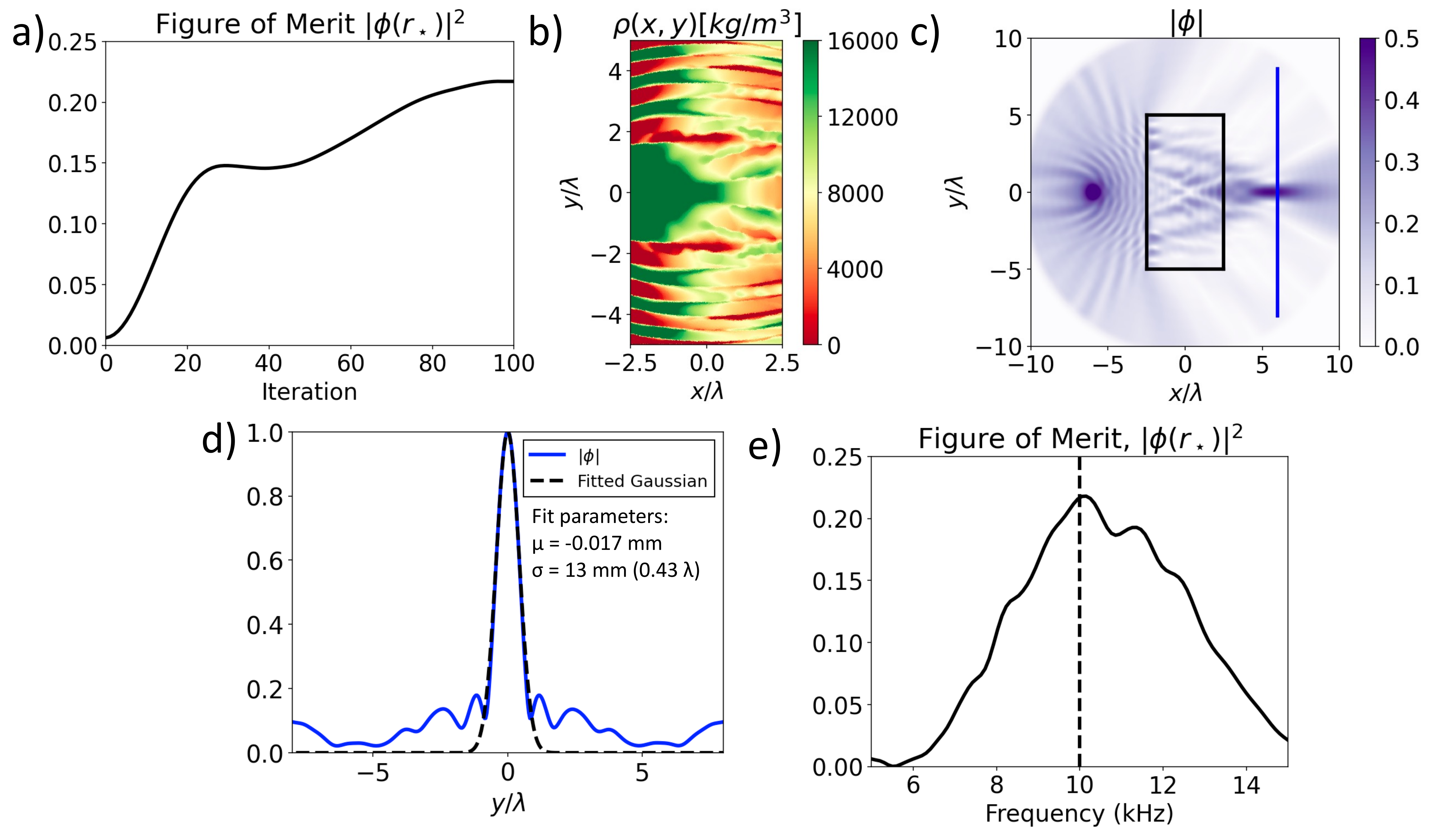}
    \caption{A graded density lens, designed using the adjoint method.
    The change in the figure of merit over the optimisation is shown in a), with the final mass density shown in b).  
    For reference, the initial mass density was that of steel $\sim 8000$kg/m$^3$.
    Displacement amplitude of the final device is shown in c).
    d) Plotting the field along the cut line indicated by a blue line in c), we find that the width of the focus is $0.43 \lambda$.
    The performance of the lens as a function of frequency e) indicates a full--width half maximum bandwidth of $\sim$ 5 kHz.}
    \label{fig:lens}
\end{figure*}

As well as mass--loading or pinning, the infinite mass limit we considered in the previous section, the mechanical properties of the plate itself can alternatively be continuously graded in space \cite{Jin2019, Climente2014}.
In this section, we take inspiration from the adjoint method in electromagnetism \cite{Lalau-Keraly2013} to develop a simple and intuitive framework to design graded plates for a range of applications.
While the adjoint method was first developed in structural mechanics \cite{Bendsoe2003}, it has not been extensively used to design graded elastic systems for wave manipulation.
Returning to the wave equation for time--harmonic elastic waves on thin plates
\begin{equation}
    \left( \nabla^4 - \frac{h \omega^2}{D} \rho \right) \phi (\boldsymbol{r}) = 0 ,
    \label{eq:wave_eqn_graded}
\end{equation}
we now consider the effect of grading the properties of the plate in space.
For ease of implementation our optimisation grades the mass density $\rho$ in space, although through the dispersion relation this is then converted to a plate height grading.
Asking how a small change in mass density produces a small change in the fields, we perturbatively expand the mass density $\rho = \rho + \delta \rho$ and the fields $\phi = \phi + \delta \phi$.
To first order we find that
\begin{equation}
    \left( \nabla^4 - \frac{h \omega^2}{D} \rho \right) \delta \phi = \frac{h \omega^2}{D} \phi \delta \rho.
\end{equation}
This inhomogeneous equation can be solved by integrating the source term against the Green's function \cite{morse1953}
\begin{equation}
    \delta \phi (\boldsymbol{r}) = \frac{h \omega^2}{D} \int G(\boldsymbol{r}, \boldsymbol{r'}) \delta \rho (\boldsymbol{r'}) \phi (\boldsymbol{r'}) d^2 \boldsymbol{r'}.
\end{equation}
Assuming that the mass density changes at a single point $\boldsymbol{r}_i$ by an amount $\Delta \rho$, so that $\delta \rho (\boldsymbol{r}) = \Delta \rho \delta (\boldsymbol{r}-\boldsymbol{r}_i)$, the integral can be easily evaluated to give
\begin{equation}
    \delta \phi (\boldsymbol{r}) = \frac{h \omega^2}{D} G(\boldsymbol{r}, \boldsymbol{r}_i) \phi (\boldsymbol{r}_i) \Delta \rho .
\end{equation}
Just as Eq. \ref{eq:fieldChange} allowed us to connect a small change in the position of a pinned point to a change in the fields, this expression allows us to connect a change in the mass density to a change in the fields.
As before, we will utilise this to efficiently calculate derivatives of figures of merit.
Beginning with the same example as the previous section, we see to design a graded structure that maximises $\mathcal{F} = |\phi (\boldsymbol{r}_\star)|^2$.
Expanding under small changes in the field caused by mass density variations, we find that
\begin{align}
    \delta \mathcal{F} &= 2{\rm Re} \left[ \phi^* (\boldsymbol{r}_\star) \delta \phi (\boldsymbol{r}_\star) \right] \\
    &= \frac{2h \omega^2}{D} {\rm Re} \left[ \phi^* (\boldsymbol{r}_\star) G(\boldsymbol{r}_\star, \boldsymbol{r}_i) \phi (\boldsymbol{r}_i) \right] \Delta \rho .
    \label{eq:fieldChange2}
\end{align}
Exploiting the reciprocity of the Green's function Eq. (\ref{eq:g}) to swap the two arguments, then re--ordering, we can re-cast Eq. (\ref{eq:fieldChange2}) as
\begin{equation}
    \delta \mathcal{F} = \frac{2h \omega^2}{D} {\rm Re} \left[ \phi (\boldsymbol{r}_i) G(\boldsymbol{r}_i, \boldsymbol{r}_\star) \phi^* (\boldsymbol{r}_\star) \right] \Delta \rho .
\end{equation}
This can now be evaluated with only two field calculations.
The first is the `forwards field' $\phi (\boldsymbol{r}_i)$ which is the field due to the source evaluated at each of the locations at which we seek to grade the mass density (or height of the plate).
Second is the `adjoint field' $G(\boldsymbol{r}_i, \boldsymbol{r}_\star) \phi^* (\boldsymbol{r}_\star)$, which is the field due to a point source at the target location $\boldsymbol{r}_\star$ evaluated at each point we wish to grade the density at $\boldsymbol{r}_i$, scaled by the value of the field at the target location.
In terms of these quantities, the gradient of the figure of merit with respect to the density distribution is given by
\begin{equation}
    \pdv{\mathcal{F}}{\rho} = {\rm Re} \left[ \phi (\boldsymbol{r}_i) \phi_{\rm adjoint} (\boldsymbol{r}_i) \right] ,
    \label{eq:adjointGrad}
\end{equation}
where the adjoint field is defined as
\begin{equation}
    \phi_{\rm adjoint} (\boldsymbol{r}) = \frac{2h \omega^2}{D} G(\boldsymbol{r}, \boldsymbol{r}_\star) \phi^* (\boldsymbol{r}_\star) .
    \label{eq:adjoint_field}
\end{equation}
Calculation of the displacement of the plate with a graded mass density is facilitate using finite element simulations in COMSOL Multiphysics \cite{comsol}.
An illustration of how the adjoint method works is shown in Figure \ref{fig:adjoint_method}.
Seeking to design a graded density lens for flexural waves, we place a point source at $(-6\lambda, 0)$, generating a field shown Figure \ref{fig:adjoint_method} a).
This is our source or `forwards' field.
Choosing the focus of our lens to be at $(6\lambda, 0)$, we place the adjoint source here, calculating the adjoint field according to Eq. (\ref{eq:adjoint_field}), shown in Figure \ref{fig:adjoint_method} b).
Multiplying these two fields then taking the real part, according to Eq. (\ref{eq:adjointGrad}), we find the gradient of the figure of merit shown in Figure \ref{fig:adjoint_method} c).
The region we will grade is indicated as a black rectangle.
Green regions indicate where the mass density should be increased to increase the figure of merit, while red regions indicate where the mass density should be decreased.
An example of using this framework to design a graded lens for waves on elastic plates is shown in Figure \ref{fig:lens}.
The progression of the figure of merit is shown in Figure \ref{fig:lens} a), with the final structure shown in b) and the displacement amplitude of the device in d).
Considering a slice of the field through the focus, shown in Figure \ref{fig:lens} d), we find that the focal spot has a width of 0.43 $\lambda$.
It can be noted that the final structure shown in Figure \ref{fig:lens} b) does not exactly correspond to the gradient shown in Figure \ref{fig:adjoint_method} c).
As the optimisation progresses, the structure changes leading to the forwards field $\phi (\boldsymbol{r}_i)$ in the gradient Eq. (\ref{eq:adjointGrad}) changing.
Lastly, Figure \ref{fig:lens} e) shows the bandwidth of the lens.
Peak performance is at 10 kHz, the design frequency, although the device functions well over a much larger band than the devices based on multiple scattering of pinned points in the previous section.

\begin{figure}[h!]
    \centering
    \includegraphics[width=\linewidth]{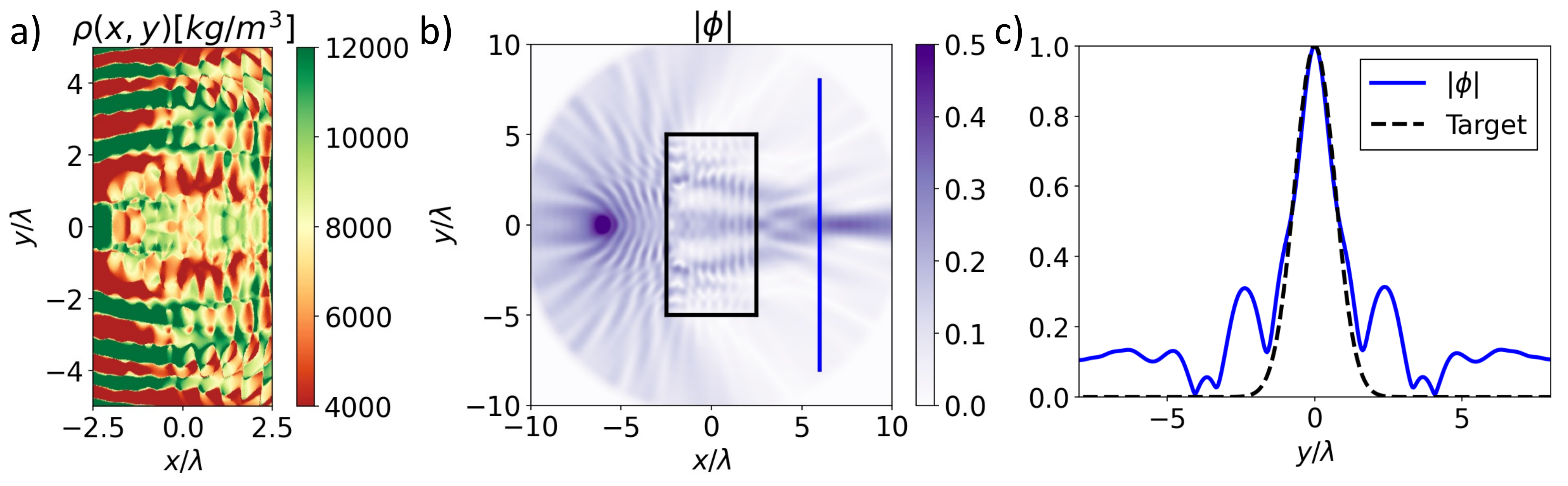}
    \caption{Mode shaping with a graded plate.
    Aiming to create a particular field distribution, the adjoint method is applied to design the density grading shown in a).
    The displacement amplitude of the device is shown in b), with a cut of the field along the line the amplitude has been designed on shown in c).}
    \label{fig:gaussian_wide}
\end{figure}

\begin{figure}[h!]
    \centering
    \includegraphics[width=\linewidth]{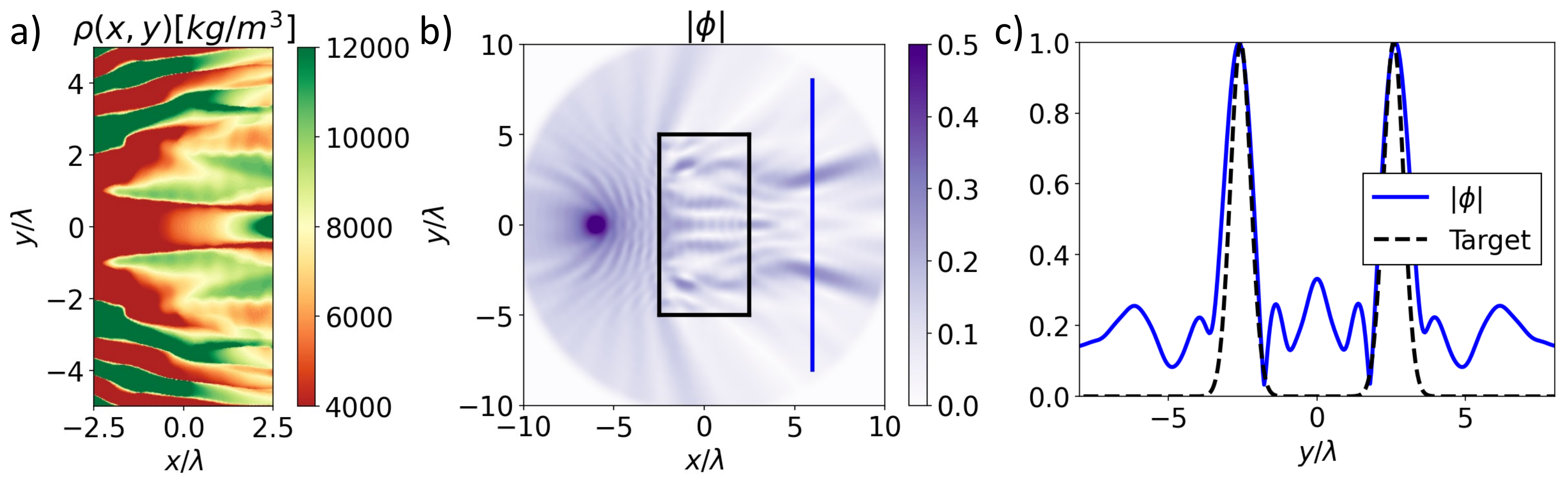}
    \caption{Graded plate with a designed displacement amplitude distribution along $x = 6\lambda$.
    The density grading is shown in a), with the displacement amplitude shown in b) and a plot along the $x = 6\lambda$ line shown in c).}
    \label{fig:gaussian_double}
\end{figure}

The adjoint method can also be leveraged to choose the exact distribution of the field in space.
Say we have a 1D line in space on which we would like to choose the shape of the displacement amplitude.
The figure of merit we must optimise is 
\begin{equation}
    \mathcal{F} = \int dy |\phi(y)| T(y) ,
    \label{eq:modeShapingFoM}
\end{equation}
where $T(y)$ is the target displacement amplitude distribution.
As before, we aim to find an analytic expression for the gradient of this figure of merit in terms of a forwards field and an adjoint field.
To this end, we expand the modulus of the displacement field under small perturbations
\begin{align}
    |\phi + \delta \phi| &= \sqrt{(\phi^* + \delta \phi^*)(\phi + \delta \phi)} , \\
    &= |\phi|\sqrt{1 + \frac{2{\rm Re}[\phi^* \delta \phi]}{|\phi|^2}} , \\
    &\approx |\phi| + \frac{1}{|\phi|} {\rm Re}[\phi^* \delta \phi] ,
\end{align}
finding that the first order correction is
\begin{equation}
    \delta |\phi(y)| = \frac{1}{|\phi (y)|} {\rm Re}[\phi^* (y) \delta \phi (y)] .
\end{equation}
Substituting this back into the figure of merit Eq. \ref{eq:modeShapingFoM}, the first order change to the figure of merit is
\begin{align}
    \delta \mathcal{F} &= {\rm Re} \left[ \int dy \frac{1}{|\phi (y)|} T(y) \phi^* (y) \delta \phi (y) \right] , \\
    &= \frac{h \omega^2}{D} {\rm Re} \left[ \int dy \frac{1}{|\phi (y)|} \phi^* (y) G(y, \boldsymbol{r}_i) \phi(\boldsymbol{r}_i) T(y) \right] \Delta \rho .
\end{align}
Exploiting reciprocity to swap the arguments of the Green's function, we can identify the adjoint field as 
\begin{equation}
    \phi_{\rm adjoint} (\boldsymbol{r}) = \frac{h \omega^2}{D} \int dy \frac{\phi^* (y)}{|\phi (y)|} G(\boldsymbol{r}, y) T(y) ,
    \label{eq:adjointShaping}
\end{equation}
so that the gradient of the figure of merit is given by $\partial \mathcal{F} / \partial \rho = {\rm Re} [\phi (\boldsymbol{r}) \times \phi_{\rm adjoint} (\boldsymbol{r})]$.
At this point, we note that the adjoint field has a simple physical interpretation.
The solution to a inhomogeneous differential equation can be found by integrating the source term against the Green's function.
We can therefore conclude, based on the form of Eq. (\ref{eq:adjointShaping}), that the adjoint field is generated by a distributed source that is the mode distribution we aim for $T(y)$, up to a phase factor given by $\phi^* (y)/|\phi(y)|$.
This has been noted before in the design of electromagnetic waveguide systems \cite{Lalau-Keraly2013}.
To shape the field in space one must set up the adjoint source to be the desired field distribution then solve the wave equation (\ref{eq:wave_eqn_graded}) to find the adjoint field.
This is then interfered with the forwards field to find how to grade the plate.
An example of this is shown in Figure \ref{fig:gaussian_wide}, where the target displacement amplitude distribution at $x = 6\lambda$ was chosen to be 
\begin{equation}
    T(y) = \exp \left[-\frac{1}{2}\left( \frac{y}{0.02}\right)^2 \right].
\end{equation}
The resulting density grading is shown in Figure \ref{fig:gaussian_wide} a), with the field distribution shown in b).
A cut of the field along with the target distribution is shown in Figure \ref{fig:gaussian_wide} c).
As another example, we choose the target distribution of the field to be
\begin{equation}
    T(y) = \exp \left[-\frac{1}{2}\left( \frac{y-0.08}{0.015}\right)^2 \right] + \exp \left[-\frac{1}{2}\left( \frac{y+0.08}{0.015}\right)^2 \right].
\end{equation}
Using this as the adjoint source, the resulting device is shown in Figure \ref{fig:gaussian_double}.

\section{Conclusions \& Outlook}

Motivated by the recent development of simple but versatile methods for designing electromagnetic materials, we have presented two perturbative methods for designing thin elastic plates for a range of applications.
While flexural waves on thin elastic plates have been studied for many years, the question of how to structure plates to achieve various wave shaping effects remains open.
Here, we develop an iterative method for deciding where to pin a thin plate for focusing mechanical energy or isolating regions of space.
Next, we apply the adjoint method to thin elastic plates, allowing graded density of height profiles to be found that focus energy or enable mode shaping.
Both methods are applicable to a range of problems, from energy harvesting to vibration isolation.
We have also noted that the devices based on multiple scattering effects between many pinned points exhibit lower bandwidth than graded devices.
This is due to the sensitivity of multiple scattering effects to changes in scatterer positions.

The methods we have presented are easily extendable to vector elasticity, where one could try to affect mode conversion or design materials that manipulate different polarisations in different ways.
More exotic wave--scattering effects could also be achieved by building up a system of resonant elements with unusual properties, such as ones with negative effective density or stiffness.

\section*{Acknowledgements}
    J.R.C. would like to thank Dr G.J. Chaplain and Prof S.A.R. Horsley for many useful discussions and for their feedback on the manuscript.
    
    We acknowledge financial support from the Engineering and Physical Sciences Research Council (EPSRC) of the United Kingdom, via the EPSRC Centre for Doctoral Training in Metamaterials (Grant No. EP/L015331/1) and from Defence Science Technology Laboratory (DSTL).

\section*{Data Availability}
    All data and code created during this research are openly available at \url{https://doi.org/10.24378/exe.XXXX}.


\end{document}